\documentclass{emulateapj}
\usepackage{apjfonts}
\usepackage{natbib}
\usepackage{graphicx}

\begin{document}
\title{Discovery of VHE Gamma Radiation from IC~443 with
the MAGIC Telescope}

\newcommand{\commentOut}[1]{}
\newcommand{\commentOn}[2]{{\emph #1}{\em {(\footnotesize Comment: #2)}}}

\author{
 J.~Albert\altaffilmark{a}, 
 E.~Aliu\altaffilmark{b}, 
 H.~Anderhub\altaffilmark{c}, 
 P.~Antoranz\altaffilmark{d}, 
 A.~Armada\altaffilmark{b}, 
 C.~Baixeras\altaffilmark{e}, 
 J.~A.~Barrio\altaffilmark{d},
 H.~Bartko\altaffilmark{f,}\altaffilmark{*},
 D.~Bastieri\altaffilmark{g}, 
 J.~K.~Becker\altaffilmark{h}
% ...
}
 \altaffiltext{a} {Universit\"at W\"urzburg, D-97074 W\"urzburg, Germany}
 \altaffiltext{b} {IFAE, Edifici Cn., E-08193 Bellaterra (Barcelona), Spain}
 \altaffiltext{c} {ETH Zurich, CH-8093 Switzerland}
 \altaffiltext{d} {Universidad Complutense, E-28040 Madrid, Spain}
 \altaffiltext{e} {Universitat Aut\`onoma de Barcelona, E-08193 Bellaterra, Spain}
 \altaffiltext{f} {Max-Planck-Institut f\"ur Physik, D-80805 M\"unchen, Germany}
 \altaffiltext{g} {Universit\`a di Padova and INFN, I-35131 Padova, Italy}  
 \altaffiltext{h} {Universit\"at Dortmund, D-44227 Dortmund, Germany}
% ...
\altaffiltext{*} {correspondence: H.~Bartko, hbartko@mppmu.mpg.de, C.~Delgado, delgadom@iac.es}


\begin{abstract}

We report the detection of a new source of very high energy (VHE, $E_{\gamma}\geq 100$~GeV)
$\gamma$-ray emission 
located close to the Galactic Plane, MAGIC~J0616+225,
which is spatially coincident with SNR IC~443.
The observations were carried out with the MAGIC telescope
in the periods
December 2005 - January 2006 and December 2006 - January 2007.
Here we present results from this source, leading to a 
VHE $\gamma$-ray signal with a statistical significance of
5.7 sigma in the 2006/7 data and a measured
differential $\gamma$-ray flux consistent with a
power law, described as $\mathrm{d}N_{\gamma}/(\mathrm{d}A
\mathrm{d}t \mathrm{d}E) = (1.0 \pm 0.2) \times 10^{-11}
(E/\mathrm{0.4 TeV})^{-3.1 \pm 0.3} \ \mathrm{cm}^{-2}\mathrm{s}^{-1}
\mathrm{TeV}^{-1}$. 
We briefly discuss the observational technique used and the
procedure implemented for the data analysis.
The results are 
placed in the context
of the multiwavelength emission and the 
molecular environment found in
the region of IC~443. 

\end{abstract}


\section{Introduction}

IC~443 is an asymmetric shell-type SNR with a diameter of 45 arc
minutes at a distance of about 1.5~kpc \citep{Fesen1984,Claussen1997}.
It is included in Green's catalog \citep{Green2004},
and it has a spectral index of 0.36, and a flux density of 160~Jy at 
1~GHz. It was mapped in radio with the VLA at 90~cm \citep{Claussen1997}
and at 20, 6 and 3.5~cm \citep{Olbert2001,Condon1998}.  Moreover, \citet{Claussen1997}
reported the presence of maser emission at 1720~MHz from four sources, the strongest of which is located at $(l,b)\sim(-171.0,\, 2.9)$.
IC~443 is a prominent X-ray source, with data available from Rosat \citep{AsaokaAschenbach1994}, ASCA \citep{Keohane1997}, XMM \citep{BocchinoBykov2000,BocchinoBykov2001,BocchinoBykov2003,Bykov2005,Troja2006}, and Chandra \citep{Olbert2001,Gaensler2006}.
The EGRET has detected a $\gamma$-ray source above 100 MeV in the IC~443 SNR, 3EG J0617+2238 \citep{Eposito1996,Hartman1999}.
Upper limits to the very high energy (VHE) $\gamma$-ray emission from IC~443 were reported by the Whipple collaboration: $\mathrm{d}N_{\gamma}/(\mathrm{d}A \mathrm{d}t) < 6 \times 10^{-12} \mathrm{cm}^{-2}\mathrm{s}^{-1}$ (0.11 Crab) above 500~GeV \citep{Holder2005} and by the CAT collaboration $\mathrm{d}N_{\gamma} / (\mathrm{d}A \mathrm{d}t) < 9 \times 10^{-12} \mathrm{cm}^{-2}\mathrm{s}^{-1}$ above 250~GeV \citep{Khelifi2003}.

Here we present observations of the SNR IC~443 with the Major Atmospheric
Gamma Imaging Cherenkov (MAGIC) telescope
resulting in the detection of a new source of VHE
 $\gamma$-rays, named MAGIC~J0616+225. We briefly
discuss the observational technique used and the procedure
implemented for the data analysis, derive a VHE
$\gamma$-ray spectrum, and analyze it in comparison
with other observations, including the molecular environment found
in the region of IC~443.

\newpage

\section{Observations}

MAGIC (see e.g., \citet{MAGIC-commissioning,CortinaICRC} for a
detailed description) is the largest single dish Imaging Air
Cherenkov Telescope (IACT) in operation\footnote{
see also http://wwwmagic.mppmu.mpg.de/magic/factsheet/}. It is located on the Canary Island 
La Palma ($28.8^\circ$N, $17.8^\circ$W, 2200~m a.s.l.).

The SNR IC~443 was observed for a total of 10 hours in the period December 2005 - January 2006 (period I), with the telescope pointing to the SNR center. 
The VHE $\gamma$-ray sky map showed 
evidence for a VHE $\gamma$-ray signal
with a statistical significance of 3$\sigma$ (before trials were taken into account). 
To test the hypothesis that this excess is due to a VHE $\gamma$-ray source, MAGIC~J0616+225, the excess center was observed for a total of 37 hours in
the period December 2006 - January 2007 (period II). Therefore, in the analysis of the period II data no trial factors needed to be taken into account to compute the statistical significance of the source detection.
Changes in the readout chain in spring 2006
due to the installation of a novel 2~GSamples/s FADC system \citep{MUX_tests, MUX_ICRC}
in parallel to the existing 300~MSamples/s FADC system 
made it advisable (on grounds of simplicity of the analysis) to select only the second part of the dataset, period II, 
for the 
studies presented here. However, within statistics the data from period I provide 
compatible results.

At La Palma, IC~443 culminates at about $6^\circ$ zenith angle
(ZA). 
%
%
The 
observations 
were carried out in the false-source tracking
(wobble) mode \citep{wobble}. The sky directions (W1, W2) to be
tracked were on two opposite sides of the source direction, at a distance of 
$0.4^\circ$ from the source. They were chosen
such that in the camera the sky brightness distribution
relative to W1 was similar to the one relative to W2.
For each
tracking position three circular background control regions were defined, 
located symmetrically to the source region 
with respect to the
camera center. During wobble mode data taking, 50\% of
the data was taken at W1 and 50\% at W2, switching (wobbling)
between the 2 directions every 20 minutes. 
%
%
%
%
In total, about 8 million triggers were recorded in period I
and about 30 million triggers were recorded in 
period II.
There are two bright stars in the field of view: $\eta$~Gem ($\mathrm{mag(V)}=3.28$ and $\mathrm{B}-\mathrm{V} = 1.60$) at a distance to MAGIC J0616+225 of $0.4^{\circ}$ and $\mu$~Gem ($\mathrm{mag(V)}=2.88$ and $\mathrm{B}-\mathrm{V} = 1.64$) at a distance of $1.9^{\circ}$ to MAGIC J0616+225. Both stars are rather red such that the increase in the pixel anode currents is still sustainable at nominal pixel high voltage \citep{MAGIC_moon,Paneque2004}. The discriminator thresholds of the channels included in the trigger are dynamically regulated to keep the individual pixel rates at a constant level \citep{CortinaICRC}.


\section{Data Analysis}


\begin{figure}[t]
\begin{center}
\includegraphics[width=\columnwidth]{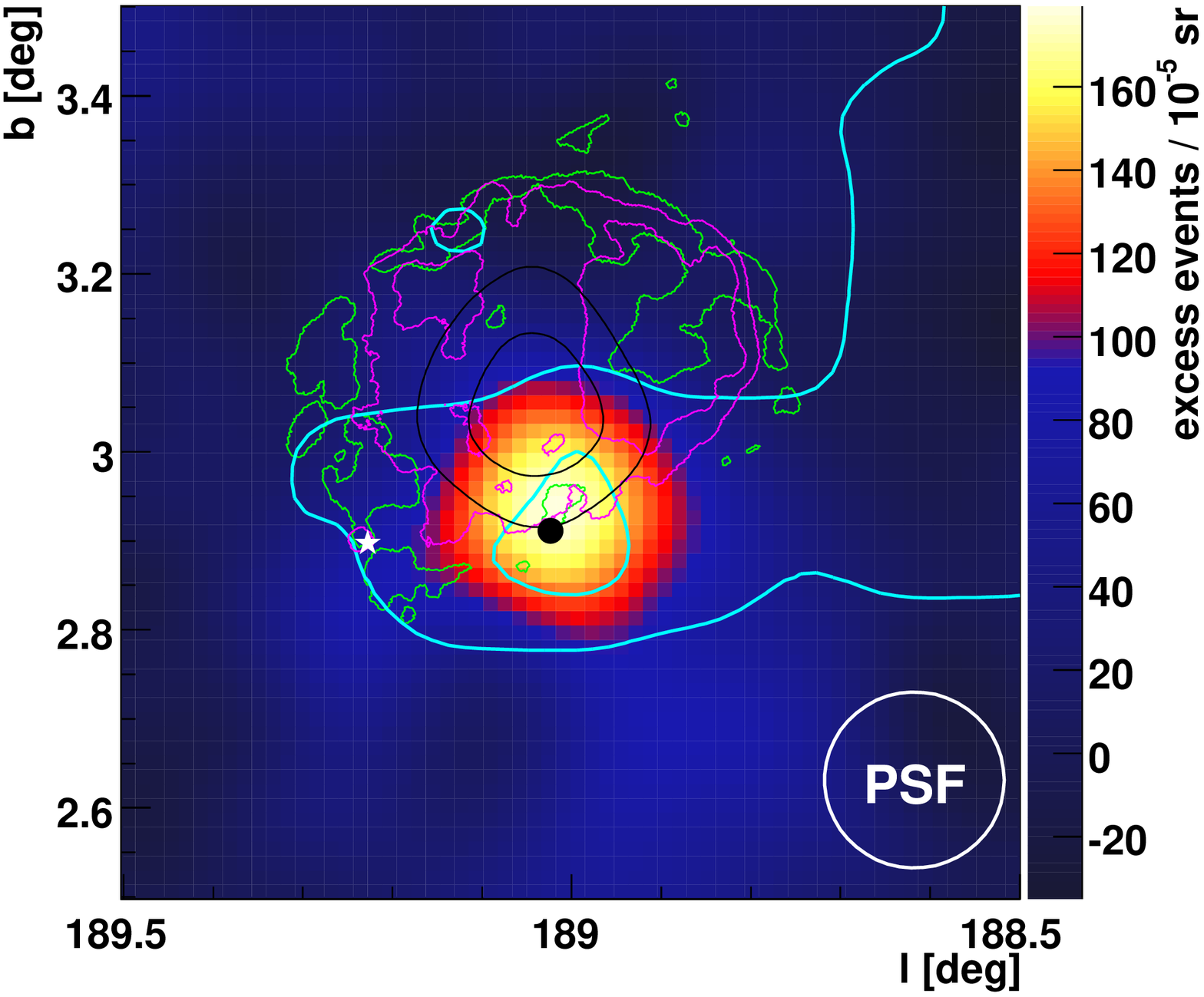}
\end{center}
\caption{Sky map of $\gamma$-ray candidate events (background
subtracted) in the direction of MAGIC~J0616+225 for an energy threshold
of about 150~GeV in galactic coordinates. 
Overlayed are $^{12}$CO  emission contours
(cyan) from \citet{Dame2001}, contours of 20 cm VLA radio data
from \citet{Condon1998} (green), X-ray contours from Rosat \citep{AsaokaAschenbach1994} (purple) and $\gamma$-ray contours from EGRET \citep{Hartman1999} (black).
{The $^{12}$CO contours are at
7 and 14 K km/s, integrated from -20 to 20 km/s in velocity, the
range that best defines the molecular cloud associated with IC~443.
The contours of the radio emission are at
5 mJy/beam, chosen for best showing
both the SNR IC~443. 
The X-ray contours are at 700 and 1200 counts / $6 \cdot 10^{-7}$ sr. The EGRET contours represent a 68\% and 95\% statistical probability that a single source lies within the given contour.
The white star denotes the position of the pulsar CXOU J061705.3+222127 \citep{Olbert2001}. The black dot shows the position of the 1720 MHz OH maser \citep{Claussen1997}.} The white circle shows the MAGIC PSF of $\sigma = 0.1^{\circ}$.
}
\label{fig:disp_map}
\end{figure}


\begin{figure}[!h]
\begin{center}
\includegraphics[totalheight=5cm]{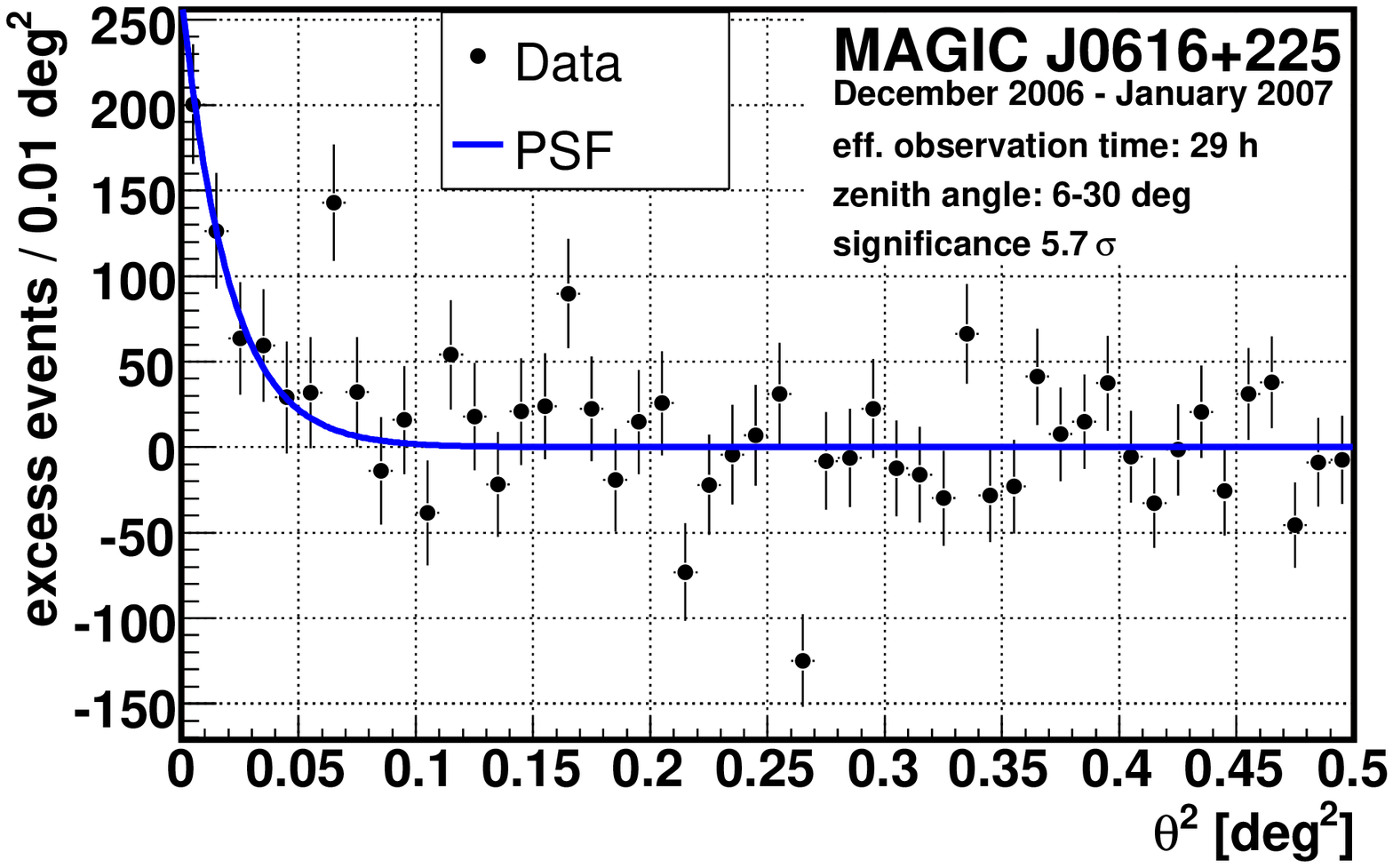}
\end{center}
\caption{Distribution of excess $\gamma$-ray candidate events (figure \ref{fig:disp_map}) as a function of the squared angular distance from the excess center of MAGIC~J0616+225 (points), compared to the expected distribution for a point-like source (blue line) corresponding to the MAGIC PSF.
} \label{fig:theta2}
\end{figure}


%


\begin{figure}[!h]
\begin{center}
\includegraphics[totalheight=5cm]{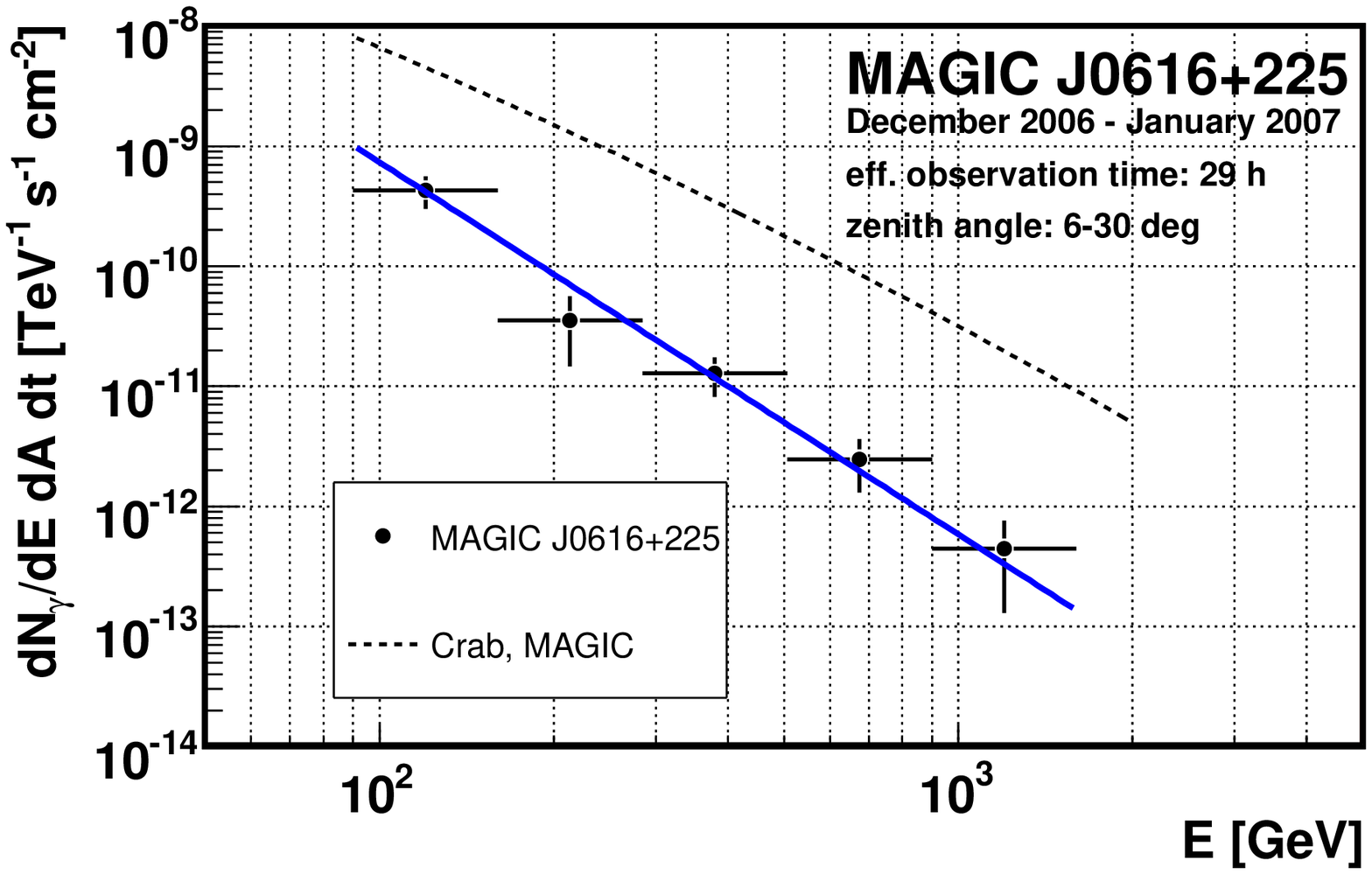}
\end{center}
\caption{VHE $\gamma$-ray spectrum of MAGIC~J0616+225 (statistical errors only).
The full line shows the result of a simple power law fit to the 
spectral points taking into account correlations between the spectral points that are introduced by the unfolding procedure.
The dashed line shows the spectrum of the Crab Nebula as measured by MAGIC
\citep{Crab_MAGIC}.
} \label{fig:spectrum}
\end{figure}


Data runs with anomalous trigger rates due to bad observation conditions (three nights of data taking) were rejected from further analysis. In addition, data with ZA$> 30^{\circ}$ (2~h) were excluded.
The remaining data set corresponded to an effective observation time (including 
a dead-time correction)
of 29 h.

The data analysis was carried out using the standard MAGIC
analysis and reconstruction software \citep{Magic-software}, the
first step of which involves the FADC signal reconstruction using
a digital filter and the calibration of the raw data
\citep{MAGIC_signal_reco,MAGIC_calibration}. 
Thereafter,
shower images were cleaned by applying a cut of 10 photoelectrons (ph.~el.)
for image core pixels and 5 ph.~el. for boundary pixels, respectively 
(see e.g. \citep{Fegan1997}). These tail
cuts were 
scaled for the larger size of the outer
pixels of the MAGIC camera. The camera images were characterized by
image parameters \citep{Hillas_parameters}. In this analysis, the
Random Forest method (see \citet{RF,Breiman2001} for a detailed
description) was applied for the $\gamma$/hadron separation (for a
review see e.g. \citet{Fegan1997}) as well as for the energy estimation, see e.g. \citet{MAGIC_1813,MAGIC_GC,MAGIC_1834,Crab_MAGIC}.
Above the 
analysis energy threshold,
the cut efficiency reaches about 50\% 
corresponding to an effective collection area for $\gamma$-ray 
showers of about 50000 m$^2$. 


For each event the arrival direction of the primary $\gamma$-ray
candidate in sky coordinates was estimated using the DISP-method
\citep{wobble,Lessard2001,MAGIC_disp}. A conservative 
cut of SIZE$\geq 200$~ph.~el. was applied to select a subset of events with
superior angular resolution. For these events the effect of the bright stars in the field of view is negligible. The corresponding analysis energy
threshold (the energy corresponding to the maximum of the differential
$\gamma$-ray rate after all analysis cuts 
) was about 150~GeV.

Figure \ref{fig:disp_map} shows the sky map of $\gamma$-ray
candidates (background subtracted) 
from the direction of MAGIC~J0616+225 in galactic coordinates. 
It was folded with a
two-dimensional Gaussian with a standard deviation of
$0.072^{\circ}$.
The MAGIC $\gamma$-ray PSF (standard deviation of a two-dimensional 
Gaussian fit to the non-folded brightness profile of a point source) 
is $(0.10\pm0.01)^{\circ}$ for SIZE$\geq 200$~ph.el. \citep{MAGIC_disp}. The folding of the sky map serves
to increase the signal-to-noise ratio by smoothing out
statistical fluctuations. However, it correlates the values of neighbouring bins 
and somewhat degrades
the spatial resolution. The sky map is overlayed with contours
of $^{12}$CO  emission
(cyan) from \citet{Dame2001}, contours of 20 cm VLA radio data
from \citet{Condon1998} (green), X-ray contours from Rosat (purple) \citep{AsaokaAschenbach1994} and $\gamma$-ray contours from EGRET \citep{Hartman1999} (black).
Also XMM-Newton and Chandra X-ray observations of IC~443 exist; they will be discussed below.
The white star denotes the position of the pulsar CXOU J061705.3+222127 \citep{Olbert2001}. The black dot shows the position of the 1720 MHz OH maser \citep{Claussen1997}.
%
The VHE $\gamma$-ray sky map shows a clear excess centered at (RA,
DEC)=(06$^\mathrm{h}$16$^\mathrm{m}$43$^\mathrm{s}$,
+22$^\circ$31'48''), MAGIC~J0616+225. The statistical position error is 1.5',
the systematic error due to background determination and 
pointing uncertainty is estimated to be 1' \citep{Crab_MAGIC}. 
Within errors MAGIC~J0616+225 is point-like. 
The center of gravity of the excess in the period II data presented here agrees with the center of gravity of the excess in the period I data not shown here.

Figure \ref{fig:theta2} shows for the excess $\gamma$-ray-like events in figure \ref{fig:disp_map} the distribution of the squared
angular distance, $\theta^2$, with respect to the 
excess center together with the expected distribution for a point-like source.
Chosing a conservative $\theta^2$ cut of $\theta^2 \leq 0.05 \, \mathrm{deg}^2$, as appropriate for unidentified sources, the observed excess in the period II data set in the
direction of MAGIC~J0616+225 has a significance of 5.7$\sigma$. 
The period I data showed an excess of about 3$\sigma$ significance at a position less then $3'=0.05^{\circ}$ away from period II data set.
Computing the significance of the period II data set with respect to the period I position yields only a negligible difference for the chosen large signal region $\theta^2 \leq 0.05 \, \mathrm{deg}^2$.
Chosing smaller $\theta^2$ cuts, one obtains a higher significance for the period II excess center.


For the spectral analysis the excess data from a sky region of maximum angular distance
of $\theta^2 = 0.05 \, \mathrm{deg}^2$ around the excess center
were integrated. Figure \ref{fig:spectrum} shows the reconstructed
VHE $\gamma$-ray spectrum
($\mathrm{dN}_{\gamma}/(\mathrm{dE}_{\gamma} \mathrm{dA}
\mathrm{dt})$ vs. true $\mathrm{E}_{\gamma}$) of MAGIC~J0616+225 after
correcting (unfolding) for the instrumental energy resolution
\citep{Anykeev1991,Bertero1989}. The horizontal bars indicate the bin size in
energy.
A simple power law was fitted to the 
spectral points taking into account correlations between the spectral points that are introduced by the unfolding procedure.
The result is given by
($\chi^2/\mathrm{n.d.f}=1.1$):
\begin{eqnarray*}
\frac{ \mathrm{d}N_{\gamma}}{\mathrm{d}A \mathrm{d}t \mathrm{d}E}
= (1.0 \pm 0.2) \times  10^{-11}
\left(\frac{E}{\mathrm{0.4 TeV}}\right)^{-3.1 \pm 0.3}
 \mathrm{cm}^{-2}\mathrm{s}^{-1}
\mathrm{TeV}^{-1}.
\end{eqnarray*}
The quoted errors are statistical. The systematic error is
estimated to be 35\% in the flux level determination and 0.2 in
the spectral index, see also \citet{MAGIC_GC,Crab_MAGIC}. 
The integral flux of MAGIC~J0616+225 above 100~GeV is about 6.5\%
and above 300~GeV about 2.8\% 
of the Crab Nebula flux.
The integral flux of MAGIC~J0616+225 in the observation period II 
is compatible 
within errors with that of period I. 
Within the two observation periods
(two months each) no flux variations exceeding the
measurement errors were observed.
All data analysis steps were cross-checked by a second, independent analysis, yielding compatible results.

\section{Discussion and concluding remarks}

If, due to the spatial association shown in Figure \ref{fig:disp_map}, 
we accept that
MAGIC~J0616+225 is associated with SNR IC~443, located at a distance of
$\sim 1.5$ kpc, then it has a luminosity between 100~GeV and
1~TeV of about $ 2.7 \times 10^{33}$ erg s$^{-1}$.
This is roughly one order of magnitude below the luminosity of
the sources HESS J1813-178 and J1834-087 \citep{Aharonian2005b}, both also detected by MAGIC \citep{MAGIC_1813,MAGIC_1834}.
In addition, the $\gamma$-ray spectrum of
MAGIC~J0616+225 is steeper than those of J1813-178 and J1834-087.

Figure \ref{fig:disp_map} indeed shows a very interesting multi-frequency 
phenomenology.
First, it can be noted that the MAGIC VHE $\gamma$-ray source is slightly displaced
with respect to the central position of the EGRET source 3EG~J0617+2238, 
although still consistent with it, within errors. 
An independent analysis of GeV photons measured by EGRET resulted
in the source GeV~J0617+2237 \citep{Lamb1997},
that is at
the same location of 3EG~J0617+2238. 
\citet{Gaisser1998,Baring1999} and \citet{Kaul2001} extrapolate the energy spectrum of 3EG~J0617+2238 into the VHE $\gamma$-ray range. The predicted fluxes are higher with a harder energy spectrum than MAGIC~J0616+225. This supports the view that a direct extrapolation of the EGRET source to the MAGIC energy range is not valid.

The EGRET source is located in the center of the 
SNR, whereas the VHE $\gamma$-ray source is
displaced to the
south, in direct correlation with a molecular cloud.
The molecular cloud environment surrounding IC~443 and its possible
connection with the EGRET $\gamma$-ray source was studied by \citet{Torres2002},
who also provided a review of previous measurements
regarding this SNR environment.
There is a large amount of molecular mass ($\lesssim 10^{4}$
M$_{\odot}$) consistent with the distance to the SNR IC~443,
corresponding to a velocity range of $-20$ to 20 km/s, as shown in
Figure \ref{fig:disp_map}. The highest CO intensity detected is directly 
superimposed on
the central position of the MAGIC source. Moreover, \citet{Claussen1997}
reported the presence of maser emission from $(l,b)\sim(-171.0,
\, 2.9)$, spatially correlated with the MAGIC source (see also \citet{Hewitt2006}).
The maser emission is an indication for a shock in a high matter density environment. It is assumed to be due to collisionally excited H$_2$ molecules heated by the shock.

An electronic bremsstrahlung hypothesis for the origin
of the EGRET source
(e.g., \citet{BykovEtAl2000}) is difficult to reconcile with the fact
that the radio
synchrotron, X-rays, and optical emission is concentrated towards the
rim of the remnant, whereas
the EGRET source is located in the center.
On the other hand, the optical
emission seems to fade in regions where CO emission increases and where the
MAGIC source is located (see \citet{Lasker1990}). This
perhaps indicates that the molecular material, which absorbs
the optical radiation, is at the remnant's nearest side.
The report by \citet{Cornett1977} also argues that the molecular mass is
located between us and the SNR. Similarly, the recent analysis of XMM
observations by \citet{Troja2006} reached the same conclusions.
The observed VHE $\gamma$-radiation may be due to $\pi^0$-decays from 
interactions between cosmic rays accelerated in IC~443 and the dense 
molecular cloud.
A possible distance of this cloud from IC~443
could explain the steepness of the VHE $\gamma$-ray spectrum measured.
As has been emphasized by \citet{Aharonian1996}, the
observed $\gamma$-rays 
can have a significantly different
spectrum from that expected from the particle population at the
source (the SNR shock). 

The positions of 3EG~J0617+2238, GeV~J0617+2237, 
and MAGIC~J0616+225 are all different from that of 
the pulsar wind nebula (PWN) CXOU~J061705.3+222127 \citep{Olbert2001,BocchinoBykov2001}. The PWN is now co-located
with a high density molecular material region \citep{Seta1998}, which in addition 
is excited, as measured by a high $\mathrm{CO}(J=2-1)/\mathrm{CO}(J=1-0)$ ratio. If
the VHE $\gamma$-ray emission were related to the PWN, one may expect 
some spatial overlap between the PWN and the $\gamma$-ray sources.



A complete coverage of the X-ray
emission from the region was made with XMM \citep{BocchinoBykov2003},
resulting in the detection of 12 X-ray sources with fluxes larger than
$5\times 10^{-14}$
erg cm$^{-2}$  s$^{-1}$. None of these sources is spatially coincident
with the
MAGIC detection reported here. Rather, they are mostly
located in the
relatively small, 15 arcmin $\times$ 15 arcmin region, for which the
analysis of the 2MASS data reveals strong 2.2 $\mu$m emission
indicating interaction with
a molecular cloud. The MAGIC source, uncorrelated with X-ray sources,
is, however, also co-spatial with a region of high 2.2 $\mu$m emission,
but farther away from the shock.

In summary, the observation of IC~443 using the MAGIC Telescope has
led to the discovery of a new source of VHE $\gamma$-rays, MAGIC~J0616+225,
near the Galactic Plane.  A reasonably large data set was collected
and the spectrum of this source was measured up to energies of 1.6~TeV.
The differential energy spectrum can be fitted with a power
law of slope $\Gamma=-3.1\pm 0.3$. The coincidence
of the VHE $\gamma$-ray source with SNR IC~443 poses this
SNR as a natural counterpart, and although the mechanism
responsible for the high energy radiation remains yet to be
clarified, a massive molecular cloud and OH maser emissions are located at
the same sky position as that of MAGIC~J0616+225, and suggest that a
nucleonic origin
of the VHE $\gamma$-rays is possible.


We would
like to thank the IAC for the excellent working conditions at the
Observatory de los Muchachos in La Palma. The support of the
German BMBF and MPG, the Italian INFN and the Spanish CICYT is
gratefully acknowledged. This work was also supported by ETH
Research Grant TH~34/04~3 and the Polish MNiI Grant 1P03D01028.
%

\end{document}